# TO THE THOMPSON CROSS SECTION OF LIGHT SCATTERED BY MOVING PARTICLE

E.G.Bessonov, M.V.Gorbunkov, A.V.Vinogradov, Yu.Ya.Maslova, (*Lebedev Physical Institute, RAS*), A.A.Mikhailchenko (*Cornell University, USA*)

*Abstract.* In this paper in a framework of classical electrodynamics, we re-derived in a simple way the formula for the light scattering by moving particle with arbitrary angle of collision.

As it is well known a charged particle radiates in external fields with intensity defined by Larmor formula [1]

$$I = \frac{2e^4\gamma^2}{3m^2c^3}\left\{\left(\vec{E}+(\vec{\beta}\times\vec{H})\right)^2 - \left(\vec{\beta}\times\vec{E}\right)^2\right\}, \qquad (1)$$

where $e, m$ – are the charge and mass of particle, $\vec{\beta} = \vec{v}/c$, $\vec{v}$ - is its velocity, $\gamma = 1/\sqrt{1-\beta^2}$ -is a relativistic factor, $\beta = |\vec{\beta}|$, $c$ – is a speed of light, $\vec{E}$, $\vec{H}$ - are the vectors describing external electric and magnetic fields at the instant location of particle. The time dependence of velocity and coordinates of particle in (1) defined by variation of fields in space and time.

Below we are considering the case when the external fields are the ones associated with plane monochromatic laser wave. In this case the vector $\vec{H} = \vec{n}_L \times \vec{E}$, $(\vec{n}_L \cdot \vec{E}) = 0$, and intensity (1) comes to

$$I = \frac{c\gamma^2\sigma_{T0}}{4\pi}|\vec{E}|^2 [1-(\vec{\beta}\vec{n}_L)]^2, \qquad (2)$$

where $\vec{n}_L$ - is an unit vector in direction of propagation of wave, $\sigma_{T0} = 8\pi r_p^2/3$ - Thomson cross section of scattering of laser wave by the particle at rest, $r_p = e^2/mc^2$ - is the classical radius of particle.

The plane monochromatic wave is a particular case of undulator, which allows usage of results describing the motion and radiation of particles in undulators (see [2] and references there). Velocity in expression for intensity (1) and (2) for trajectories in undulator could be represented as $\vec{v} = \vec{\bar{v}}_\parallel + \vec{v}_\perp$, where $\vec{\bar{v}}_\parallel$ - is an average longitudinal and $\vec{v}_\perp$ - transverse components of velocity respectively.

The components of plane monochromatic elliptically polarized wave propagating along axis $x$ could be represented as the following

$$E_y = E_{ym}\sin\varphi, \ E_z = E_{zm}\cos\varphi, \ E_x = 0, \qquad (3)$$

where $\varphi = k_L x - \omega_L t + \varphi_0$ - is the actual phase, $\varphi_0$ - is initial phase, $k_L = \omega_L/c$ - is a wave vector, $\omega_L = 2\pi c/\lambda_L$ - angular frequency, $\lambda_L$ - is the wavelength. Usually the laser wave represented by a wavelet with length $l_{wp} = M\lambda_L$, where $M$ -is an integer number. Below we are considering the case when the laser wave is quasi-monochromatic ($M \gg 1$).

The frequency of oscillation of particle moving with angle in direction of wave propagation defined by expression $\omega^* = \omega_L(1-\vec{n}_L\vec{\bar{\beta}})$; the frequency of emitted radiation, in accordance with Doppler effect is equal to

$$\omega = \frac{\omega^*}{1-\vec{n}\vec{\bar{\beta}}} = \frac{\omega_L(1-\vec{n}_L\vec{\bar{\beta}})}{1-\vec{n}\vec{\bar{\beta}}}, \qquad (4)$$

where $\vec{n}$ - is unit vector in direction to observer, $\bar{\vec{\beta}} = \bar{\vec{v}}/c \cong \bar{\vec{\beta}}_\| = \bar{\vec{v}}/c$ [2]. In a weak field of laser wave of type (3), this corresponds to dipole approximation in radiation process

$$|\vec{\beta}_\perp| = |\vec{v}_\perp|/c| << 1/\gamma \qquad (5)$$

which is the same as condition $eE\lambda_L << mc^2$ (on quantum treatment-this is a condition of a single-photon scattering process), the radiation is emitted on the first harmonic only. Here $E = |\vec{E}|$. Further on we will proceed in a framework of this approximation. This yields the relations $\bar{\vec{\beta}}_\| = |\beta| = const$, $\gamma = const$.

If the particle passes the way corresponding to many oscillations in a laser field ($M >> 1$), then the emitted radiation in any direction will be monochromatic and distributed, according to (4) in a bandwidth ($\omega_{min}$, $\omega_{max}$), where

$$\omega_{min} = \frac{\omega_L(1-\vec{n}_L\bar{\vec{\beta}})}{1+|\bar{\vec{\beta}}|} \cong \frac{\omega_L(1-\vec{n}_L\bar{\vec{\beta}})}{1+\beta}, \qquad \omega_{max} \cong \frac{\omega_L(1-\vec{n}_L\bar{\vec{\beta}})}{1-\beta}. \qquad (6)$$

Spectral distribution of the energy flux, emitted by particle in a laser field under dipole radiation condition (5) defined by expression $\partial I/\partial \xi = I f(\xi)$, where $I$ -states for intensity of emitted wave (2), $f(\xi)$ - is a normalized spectral distribution of radiation over dimensionless variable $\xi = \omega/\omega_{max}$, $\int_{\xi_{min}}^{\xi_{max}} f(\xi)d\xi = 1$; here $\xi_{min} = (1-\beta)/(1+\beta)$, $\xi_{max} = 1$. Function $f(\xi)$ depends on direction of particle oscillation with respect to averaged velocity [2]. For relativistic particles ($\gamma >> 1$, $\xi_{min} << 1$) performing transverse and longitudinal harmonic oscillations these functions are

$$f_\perp(\xi) = 3\xi(1-2\xi+2\xi^2), \qquad f_\|(\xi) = 12\xi^2(1-\xi) \quad. \qquad (7)$$

In some cases the problem of radiation by moving particle could be described by quantum language in terms of the number of radiated quants. In that case the spectral distribution of photon flux of scattered laser radiation defined by expression $d\dot{N}_{ph}/d\omega = (dI/d\omega)/\hbar\omega = (dI/d\xi)/\xi\hbar\omega_{max}$, which could be represented as follows

$$\frac{d\dot{N}_{ph}}{d\xi} = \frac{I_\xi}{\hbar\xi}. \qquad (8)$$

This yields for the total photon flux

$$\dot{N}_{ph} = (I/\hbar\omega_{max})\int_0^1 [f(\xi)/\xi]d\xi. \qquad (9)$$

According to (9) it does not depend on choice of functions (7) corresponding to oscillations of particles in transverse and longitudinal directions with respect to theirs velocities and could be calculated as the following

$$\dot{N}_{ph} = 2I/\hbar\omega_m. \qquad (10)$$

The ratio of photon flux of scattered radiation to the density of incoming laser flux is $\dot{S}_L^{ph} = c|\vec{E}|^2/4\pi\hbar\omega_L$, i.e. the cross-section of scattering of laser photons by moving particle becomes equal

$$\sigma = \sigma_{T0}(1-\vec{n}_L\vec{\beta}) = \sigma_{T0}(1-\beta\cos\theta_{col}), \qquad (11)$$

where $\theta_{col}$ - is the angle of collision (the angle between vector $\vec{n}_L$ and $\bar{\vec{\beta}}_\|$ ).

From (11) it follows, that for collision angles $\theta_{col} = \pm \pi$, as one can expect, the cross section becomes equal to $\sigma = (1 \pm \beta)\sigma_{T0}$. In this case the light beam becomes shifted with respect to the moving particle on the distance which is $1 \pm \beta$ times bigger (smaller), than with respect to the particle at rest. This means, that in the cases considered the number of scattered photons per unit length of shifted laser radiation is an invariant (not depending on velocity of particle).

For collision of particles under angle $\theta_{col} = \pi/2$ the cross section $\sigma = \sigma_{T0}$ does not depend on the speed of motion in a transverse direction and becomes equal to cross section of laser photons by particle at rest. In this case the path length of particle inside laser bunch depends on particle velocity ($\sim \sqrt{1+\beta^2}$).

Formula for cross-sections of massive beams colliding with arbitrary angle was obtained by W.Pauli (1933) [3] and reflected in [1], [4]. From there, as a particular case, for the photon scattering by moving particle (for the photon $\vec{\beta}_{ph} = \vec{n}_L$) one can obtain the ratio of cross sections as

$$\sigma/\sigma_{T0} = \sqrt{(\vec{n}_L - \vec{\beta})^2 - (n_L \times \vec{\beta})^2} \ . \tag{12}$$

The above relation (12) could be transformed into the (11) by simple algebra.

Dependence of Thomson cross-section on velocity vector of moving particle describes total flux of scattered photons. By knowing parameters of laser beam and using relation (10) one can determine the number of photons scattered by a particle per single collision.

So deriving the cross section formula for scattering of photons by moving particle, we operate by electrodynamic formulas (see [1], [2]) not using relativistic transformations of special relativity (invariants, transformations between moving systems of reference). The final result represented in a simple form (10). Our derivation is valid for scattering of long laser wavelets ($M \gg 1$), when the radiation, emitted in selected direction is a quasi-monochromatic one and could be described by functions (7). This is due to the fact that W. Pauli considered the case when the particles in each beam have the same energy. We hope that the direct derivation of invariant cross section we made has a methodological interest.

Our days the W. Pauli's formula is in use while someone calculates processes in light sources based on backscattering of electron and ion beams. (see for example [5]-[8]). The total number of scattered laser photons in this case could be calculated by the following expression (see [1])

$$\nu = \sigma_{T0} \cdot c \cdot K \int n_L n_p dV dt \tag{13}$$

where $n_L$ и $n_p$ - are the densities of laser photons and particles respectively, $dV$ - is an element of volume where the interaction occurs, $dt$ - is a time duty of interaction. Coefficient $K = 1 - \vec{n}_L \vec{\beta} = \sqrt{(\vec{\beta}_1 - \vec{\beta}_2)^2 - (\vec{\beta}_1 \times \vec{\beta}_2)^2}$ called by kinematic factor of scattering; the values $\dot{\nu} = \partial \nu / \partial t$ and $L = \nu / \sigma_{T0}$ are called by the photon flux and luminosity of light source respectively.

For definition of spectral-angular, polarization, brightness and other characteristics of light source base on radiation of particles in external fields one should use an ordinary theory of electromagnetism. In these cases the kinematic factor appears in calculations automatically.

This work was supported in part by a Program of Fundamental research RAS "Fundamental and applied problems of photonics and physics of new optical materials" and by UNK FIAN.

**REFERENCES**

[1] L.D.Landau, E.M. Lifshits, "The Field Theory", Science, GIF-ML, Moscow 1960.

[2] E.G.Bessonov, "Undulators, Undulator Radiation, Free-Electron Lasers", Trudy FIAN, v.214, Nauka, 1993, pp. 3-119 (see. http://proceedings.lebedev.ru/214-1993/ ).

[3] W. Pauli, "Über die Intensität der Streustrahlung bewegter freier Elektronen", Helvetica Physica Acta, 6 (1933), p 279.

[4] W.Herr and B.Muratori, "Concept of Luminosity", Proceedings of CERN Accelerator School, Zeuthen 2003, CERN-2006-002 (2006).
http://cds.cern.ch/record/941318/files/p361.pdf

[5] E.Bulyak and V.Skomorokhov,"Parameters of Compton X-ray beams: Total yield and pulse duration", Physical review special topics - accelerators and beams **8**, 030703 (2005).

[6] C. Sun and Y. K. Wu, "Theoretical and Simulation Studies of Characteristics of a Compton light source", Physical review special topics - accelerators and beams 14, 044701 (2011).

[7] E.G. Bessonov, Nucl.Instrum. Meth., B309 (2013) 92-94.

[8] M.W.Krasny, "The Gamma Factory Proposal for CERN", arXiv:1511.07794.